\documentclass[9pt,twocolumn,twoside]{optica}
\setboolean{shortarticle}{false}
\setboolean{minireview}{false}

\usepackage{lipsum}       %
\usepackage{xargs}        
\usepackage{amssymb}      
\usepackage{mathtools}    
\usepackage{xr}           
\usepackage[nolist]{acronym} 
\usepackage{dblfloatfix}    
\usepackage{stackengine}  

\newacro{ADC}{analog-to-digital converter}
\newacro{AIR}{achievable information rate}
\newacro{AKNS}{Ablowitz-Kaup-Newell-Segur}
\newacro{AWG}{arbitrary waveform generator}
\newacro{AWGN}{additive white Gaussian noise}
\newacro{B2B}{back-to-back}
\newacro{BER}{bit error rate}
\newacro{BPD}{balanced photodetector}
\newacro{BPS}{blind phase search}
\newacro{DAC}{digital-to-analog converter}
\newacro{DBP}{digital back-propagation}
\newacro{DC}{direct current}
\newacro{DFT}{discrete Fourier transform}
\newacro{DP-NFDM}{dual polarization nonlinear frequency division multiplexing}
\newacro{DSO}{digital storage oscilloscope}
\newacro{DSP}{digital signal processing}
\newacro{DT}{Darboux transformation}
\newacro{ENOB}{effective number of bits}
\newacro{EDFA}{erbium-doped fiber amplifier}
\newacro{FEC}{forward error correction}
\newacro{FFT}{fast Fourier transform}
\newacro{FL}{fiber laser}
\newacro{FWHM}{full-width half-maximum}
\newacro{FWM}{four-wave mixing}
\newacro{GLM}{Gelfand-Levitan-Marchenko}
\newacro{HD-FEC}{hard-decision forward error correction}
\newacro{I}{in-phase}
\newacro{IFFT}{inverse fast Fourier transform}
\newacro{INFT}{inverse nonlinear Fourier transform}
\newacro{ISI}{inter-symbol interference}
\newacro{IST}{inverse scattering transform}
\newacro{IVP}{initial value problem}
\newacro{LO}{local oscillator}
\newacro{LPA}{lossless path-averaged}
\newacro{MS}{Manakov system}
\newacro{MAP}{maximum a posteriori}
\newacro{ML}{maximum likelihood}
\newacro{MZM}{Mach-Zehnder modulator}
\newacro{MZSP}{Manakov-Zakharov-Shabat spectral problem}
\newacro{NFDM}{nonlinear frequency division multiplexing}
\newacro{NFT}{nonlinear Fourier transform}
\newacro{NIS}{nonlinear inverse synthesis}
\newacro{NLSE}{nonlinear Schr\"odinger equation}
\newacro{NMSE}{normalized mean squared error}
\newacro{OBPF}{optical band pass filter}
\newacro{ODE}{ordinary differential equation}
\newacro{OFDM}{orthogonal frequency-division multiplexing}
\newacro{OSNR}{optical signal-to-noise ratio}
\newacro{PAPR}{peak-to-average power ratio}
\newacro{PC}{polarization controller}
\newacro{PDE}{partial differential equation}
\newacro{PMD}{polarization mode dispersion}
\newacro{PSK}{phase shift keying}
\newacro{Q}{quadrature}
\newacro{QAM}{quadrature amplitude modulation}
\newacro{QPSK}{quadrature phase shift keying}
\newacro{SMF}{single mode fiber}
\newacro{SNR}{signal-to-noise ratio}
\newacro{SPM}{self-phase modulation}
\newacro{SSFM}{split step Fourier method}
\newacro{XPM}{cross-phase modulation}
\newacro{ZSP}{Zakharov-Shabat spectral problem}

\newcommand{\scatcoef}{scattering coefficients}
\newcommand{\R}{\mathbb{R}}
\newcommand{\C}{\mathbb{C}}

\newcommand{\ttm}{\tau}                         
\newcommand{\ssp}{\ell}                         
\newcommand{\fld}[1][]{E_{#1}}            
\newcommand{\nttm}{t}                           
\newcommand{\nssp}{z}                           
\newcommand{\nfld}[1][]{q_{#1}}              

\newcommand{\dispersion}{\beta_2}               
\newcommand{\nonlinfact}{\gamma}                

\newcommand{\eig}[1][]{\lambda_{#1}}                                    
\newcommand{\nfta}[1][]{a(\lambda_{#1})}                                
\newcommand{\nftaderiv}[1][]{a'(\lambda_{#1})}                          
\newcommandx{\nftb}[2][1=, 2=]{\ifthenelse{
  \equal{#2}{}}{b_{#1}(\lambda)}{b_{#1}(\lambda_#2)}}
\newcommand{\cnft}[1][]{\ifthenelse{
  \equal{#1}{}}{Q_{c}(\lambda)}{Q_{c,#1}(\lambda_#1)}}
\newcommand{\dnft}[1][]{\ifthenelse{
  \equal{#1}{}}{Q_{d}(\lambda)}{Q_{d,#1}(\lambda_#1)}}

\newcommand{\nftaconj}[1][]{\bar{a}(\lambda_{#1})}                      
\newcommandx{\nftbconj}[2][1=, 2=]{\ifthenelse{
  \equal{#2}{}}{\bar{b}_{#1}(\lambda)}{\bar{b}_{#1}(\lambda_#2)}}

\newcommand{\jostp}[1][]{\phi^{P}_{#1}(t,\lambda)}                  
\newcommand{\jostn}[1][]{\phi^{N}_{#1}(t,\lambda)}                  
\newcommand{\jostpconj}[1][]{\bar{\phi}^{P}_{#1}(t,\lambda)}        
\newcommand{\jostnconj}[1][]{\bar{\phi}^{N}_{#1}(t,\lambda)}        

\newcommand{\vecv}{v}                           
\newcommand{\auxsol}{\bar{v}}           
\newcommand{\auxsolmat}{\Theta} 

\newcommand{\deriv}[2]{\dfrac{\partial #1}{\partial #2}}                
\newcommand{\derivtwo}[2]{\dfrac{\partial^2 #1}{\partial #2^2}}         
\newcommand{\matr}[1]{\mathbf{#1}}

\newcommand{\compactmat}{\renewcommand{\arraystretch}{1}}               

\title{Dual polarization nonlinear Fourier transform-based optical communication system}

\author[1,*]{S. Gaiarin}
\author[2,*]{A. M. Perego}
\author[1]{E. P. da Silva}
\author[1]{F. Da Ros}
\author[1]{D. Zibar}

\affil[1]{DTU Fotonik, Technical University of Denmark, Lyngby, 2800 Denmark}
\affil[2]{Aston Institute of Photonic Technologies, Aston University, Aston Express Way, Birmingham, B4 7ET UK}

\affil[*]{Corresponding authors: simga@fotonik.dtu.dk, peregoa@aston.ac.uk}

\dates{Compiled \today}

\ociscodes{}

\doi{}

\begin{abstract}

New services and applications are causing an exponential increase in internet traffic.  In a few years, current fiber optic communication system infrastructure will not be able to meet this demand because fiber nonlinearity dramatically limits the information transmission rate. Eigenvalue communication could potentially overcome these limitations. It relies on a mathematical technique called "nonlinear Fourier transform (NFT)" to exploit the "hidden" linearity of the nonlinear Schr\"odinger equation as the master model for signal propagation in an optical fiber. We present here the theoretical tools describing the NFT for the Manakov system and report on experimental transmission results for dual polarization in fiber optic eigenvalue communications. A transmission of up to 373.5 km with bit error rate less than the hard-decision forward error correction threshold has been achieved. Our results demonstrate that dual-polarization NFT can work in practice and enable an increased spectral efficiency in NFT-based communication systems, which are currently based on single polarization channels.

\end{abstract}
\setboolean{displaycopyright}{true}

\begin{document}

\maketitle
\thispagestyle{fancy}
\ifthenelse{\boolean{shortarticle}}{\abscontent}{}


\section{Introduction}
Fiber optics telecommunication is the currently established  backbone infrastructure for most of the information flow across the world~\cite{Agrell2016a}. However, the demand for an always increasing transmission rate, which for the existing channels is necessarily associated with an increment of the launched signal power  to minimize the \ac{OSNR} degradation, has been predicted to be asymptotically limited by the distortion induced by the optical fiber nonlinearity~\cite{Mitra,Essiambre}. 
It is a well-known fact that light propagation in fiber optics is governed by the \ac{NLSE}~\cite{AgrawalBook} where the nonlinearity arises due to the Kerr effect.
Nonlinearity is a problem for transmitting information with the currently used modulation formats in fiber optics communications.
Indeed, as the power is increased, the signal is more distorted by the nonlinear cross-talk, thus limiting the capability of the receiver in recovering the transmitted information. It is therefore necessary to mitigate the nonlinear effects to compensate for the distortions and to provide novel approaches for the communication over the nonlinear fiber-optic channel.
Two main paths have been followed up to now to counteract this problem: the first approach consists in mitigating the nonlinear effects through a wealth of techniques such as for instance optical phase-conjugation~\cite{Ellis:16} or digital back-propagation~\cite{Ip};
the second path, more ambitiously, aims at encoding information into the eigenmodes of the nonlinear channel, whose evolution is linear upon spatial propagation.
This second approach, originally called \emph{Eigenvalue communication}, has been proposed by Hasegawa and Nyu~\cite{Hasegawa} and it is now, with various modifications,  growing as a new paradigm in optical communications~\cite{Turitsyn2017}.

This method exploits the exact integrability of the \ac{NLSE} through the \ac{IST}~\cite{Ablowitz} as the master evolution equation of the electric field propagating in \ac{SMF}.
Integrability of the \ac{NLSE} was demonstrated by Zakharov and Shabat back in 1972~\cite{Zakharov}, who found an associated spectral problem, related to a set of ordinary linear differential equations.
Following this approach, it is possible to identify the eigenvalues, that can be considered the analogous of the frequencies in the classical Fourier transform; and the so called scattering coefficients: complex amplitudes associated to the eigenvalues.
The application of the \ac{IST} to fiber optics communications allows the use of various and flexible modulation formats \cite{Turitsyn2017}.
Due to integrability, in the lossless and noiseless limit, nonlinearity is not a detrimental factor anymore, but on the contrary, it is a constitutive element of the transmission system itself.
The parallelism between the linear Fourier transform method used to solve linear initial value problems and the \ac{IST} used to solve nonlinear ones \cite{Ablowitz}, has driven some authors to rename the \ac{IST} as \ac{NFT} \cite{Yousefi2014}, which is the name currently used in the engineering communities (see~\cite{Turitsyn2017} for a recent review including historical details).
The nonlinear Fourier spectrum of a signal consists of a set of eigenvalues and the respective associated scattering coefficients.
The eigenvalues belong either to a so-called \emph{discrete spectrum} or to a \emph{continuous spectrum}; the first describes the solitonic components of the signal, while the second is associated with dispersive waves and reduces to the classical Fourier spectrum in the limit of low power.

Communications channels based on both discrete or/and continuous spectrum modulations have been extensively studied and experimentally demonstrated up to now for the scalar (single polarization) \ac{NLSE} (see e.g.~\cite{Buelow,Prilepsky,Le,Aref3,Aref2} to cite just a few).

A series of key challenges that need to be met in order to bring \ac{NFT}-based communication to exit the labs and operate in real-world infrastructures, has been described recently~\cite{Turitsyn2017}. One of those challenges consists indeed of endowing the \emph{Eigenvalue communication} approach with polarization division multiplexing, which allows information to be encoded on both orthogonal polarization components supported by \acp{SMF}.
The description of the light propagation, accounting for its polarization dynamics can, under specific conditions that apply to modern communications fiber link, be described by the Manakov equations \cite{Wai1991a}.
In a milestone paper of nonlinear science, Manakov showed that those equations can be solved analytically by the \ac{IST} \cite{Manakov1974a}. Detailed investigations of the solutions of the Manakov equations especially concerning soliton and multisoliton dynamics in presence of noise and \ac{PMD} in optical communications, as well as their connection with optical rogue waves formation, are present in the literature \cite{yang1999multisoliton,lakoba1997perturbation,xie2002influences,chen2000manakov,horikis2004nonlinear,derevyanko2006statistics,baronio2012solutions}.

To the best of our knowledge the \ac{NFT} dual polarization problem has never been tackled at the level to demonstrate a working communication system
and only very preliminary theoretical works are present in the literature on this topic~\cite{Maruta,Goossens:17}.

In this article, we present the mathematical framework underlying the dual polarization \ac{NFT} and we show an extension of our recent results on the first experimental demonstration of a  \ac{DP-NFDM} fiber optics communication system \cite{GaiarinECOC17}. We have transmitted up to 373.5~km at the \ac{HD-FEC} \ac{BER} threshold of $3.8\times10^{-3}$, with information encoded in the \ac{QPSK} modulated scattering coefficients associated with two eigenvalues belonging to the Manakov system discrete spectrum, for both orthogonal polarization components supported by a \ac{SMF}.

The structure of the paper is the following: in section~2 we will first define the \ac{NFT} for the dual polarization case and describe the mathematical tools -the \ac{DT}- needed to generate the waveforms associated with a desired nonlinear spectrum for both field polarizations. In section~3 we will discuss the details of a \ac{DP-NFDM} system. Finally, in section~4 we present a detailed account of the experimental transmission results, followed by a discussion of the results and conclusions in section~5.


\section{Mathematical framework}\label{sec:math_framework}

\subsection{Channel model}

The evolution of the slowly varying complex-valued envelopes of the electric field propagating in a \ac{SMF} exhibiting random birefringence and whose dispersion and nonlinear lengths are much larger than the birefringence correlation length, is described by the averaged Manakov equations~\cite{Wai1991a,menyuk2006interaction}
\begin{equation}
  \left\{\begin{aligned}\label{eq:MS}
    \deriv{\fld[1]}{\ssp}&=- i\dfrac{\dispersion}{2}\derivtwo{\fld[1]}{\ttm}+i\frac{8\nonlinfact}{9}\left(|\fld[1]|^2+|\fld[2]|^2\right)\fld[1]\\
    \deriv{\fld[2]}{\ssp}&=- i\dfrac{\dispersion}{2}\derivtwo{\fld[2]}{\ttm}+i\frac{8\nonlinfact}{9}\left(|\fld[1]|^2+|\fld[2]|^2\right)\fld[2]
  \end{aligned}\right.
\end{equation}
where $\ttm$ and $\ssp$ represent the time and space coordinates, \mbox{$\fld[j]$, $j = 1,2$} are the amplitudes of the two electric field polarizations, $\dispersion$ is the dispersion coefficient and $\nonlinfact$ is the nonlinearity coefficient.

In order to remove any dependency from a specific channel, it is common to work with the normalized version of \eqref{eq:MS}.
The normalized \ac{MS}~\cite{Manakov1974a,Ablowitz2004a,Docksey2000a} is obtained by performing the change of variable
\begin{equation}\label{eq:normalization}
   \nfld[j] = \dfrac{\fld[j]}{\sqrt{P}}, \hspace{0.7cm} \nttm = \dfrac{\ttm}{T_0}, \hspace{0.7cm} \nssp = -\dfrac{\ssp}{\mathcal{L}}
\end{equation}
with $P = |\dispersion|/(\frac{8}{9}\nonlinfact T_0^2)$, $\mathcal{L} = 2 T_0^2 /|\dispersion|$ and $T_0$ is a free normalization
parameter, leading to

\begin{equation}\label{eq:NMS}
  \left\{\begin{aligned}
    i\deriv{\nfld[1]}{z}&= \derivtwo{\nfld[1]}{t}+2\left(|\nfld[1]|^2+|\nfld[1]|^2\right)\nfld[1]\\
    i\deriv{\nfld[2]}{z}&= \derivtwo{\nfld[2]}{t}+2\left(|\nfld[1]|^2+|\nfld[2]|^2\right)\nfld[2]
  \end{aligned}\right.
\end{equation}
where $z$ and $t$ represent the normalized space and time variables respectively. In this study we have considered the anomalous dispersion regime ($\dispersion<0$), since it is the one that supports solitons and corresponds to the regime of currently deployed \acp{SMF}.

In realistic systems the field amplitude is attenuated upon spatial propagation at a rate $\alpha/2$, where $\alpha$ is the attenuation coefficient of the fiber. This breaks the integrability of \eqref{eq:MS}. However it is possible to suitably redefine the fields $E_{1,2}\rightarrow~E_{1,2}e^{-(\alpha/2)\ssp}$ in such a way that they obey a lossless equation with an effective nonlinearity coefficient \begin{equation}\label{eq:LPA}
\gamma_{eff}=\gamma\left(1-e^{-\alpha L}\right)/(\alpha L)
\end{equation}
where $L$ is the length of one optical fiber span.
The evolution equation with the modified nonlinear term can be considered the leading approximation of the lossy system when we account for the periodic signal boosts due to the \acp{EDFA}.
This is the so-called \ac{LPA} model~\cite{Hasegawa:90,Le:15,Kamalian}, which is in general valid when the amplifiers spacing is smaller than the soliton period, and it has been used across all the present study.

\subsection{Direct NFT}\label{ssec:direct_nft}
In order to
compute the \ac{NFT} of a signal $\nfld[1,2](t)$ it is first necessary to associate to the
\ac{MS} \eqref{eq:NMS} a so-called spectral problem. This is known for the case of the \ac{NLSE} as the \ac{ZSP}, while for the Manakov equations we can call it the \ac{MZSP}. The \ac{MZSP} is defined by the following system of linear ordinary differential equations
\begin{equation}\label{eq:zsp}
    \deriv{\vecv}{t}=\left(\lambda \mathbf{A}+\mathbf{B}\right)\vecv
\end{equation}
being
\begin{equation*}
\compactmat
\mathbf{A} = \begin{pmatrix}
      -i & 0 & 0 \\
      0 & i  & 0 \\
      0 & 0 & i
    \end{pmatrix} \ \
\mathbf{B} = \begin{pmatrix}
      0 & q_1 & q_2\\
      -q_1^* & 0 & 0 \\
      -q_2^* & 0 & 0
    \end{pmatrix} \ \
\end{equation*}
where $\vecv$ is a solution and $\eig$ is a spectral parameter.

Assuming the vanishing boundary conditions for the signal, i.e., $|\nfld[1,2](t)|\xrightarrow{} 0$ for $t\xrightarrow{} |\infty|$, it is
possible to find a set of canonical solutions to \eqref{eq:zsp} called Jost solutions defined as~\cite{Ablowitz2004a}:

\begin{subequations}\label{eq:jost}
  \begin{align}
  \compactmat
  \jostn \rightarrow \begin{pmatrix}
                1 \\
                0 \\
                0
              \end{pmatrix}e^{-i\lambda t}; \ \
  \jostnconj\rightarrow \begin{pmatrix}
                  0 & 0 \\
                  1 & 0 \\
                  0 & 1
                \end{pmatrix}e^{i\lambda t} \ \
  t\rightarrow -\infty
  \end{align}

  \begin{align}
  \compactmat
  \jostp\rightarrow  \begin{pmatrix}
                0 & 0 \\
                1 & 0 \\
                0 & 1
              \end{pmatrix}e^{i\lambda t}; \ \
  \jostpconj\rightarrow \begin{pmatrix}
                  1 \\
                  0 \\
                  0
                \end{pmatrix}e^{-i\lambda t} \ \
  t\rightarrow +\infty~.
  \end{align}
\end{subequations}
$\{\jostp,~\jostpconj\}$
and $\{\jostn,~\jostnconj\}$ are two bases for the eigenspace associate to $\lambda$. One can
write $\jostn$ and $\jostnconj$ as a linear combination of the basis vectors $\{\jostp,~\jostpconj\}$ as

\begin{subequations}\label{eq:jost_projection}
  \begin{align}\label{eq:jost_projection_1}
  \jostn&=\jostp \nftb+ \jostpconj \nfta\\
  \jostnconj&=\jostp \nftaconj+ \jostpconj \nftbconj
  \end{align}
\end{subequations}
with coefficients $a(\lambda)$, $b(\lambda)$, $\bar{a}(\lambda)$ and $\bar{b}(\lambda)$, where $a(\lambda)$ is a scalar,
$\bar{a}(\lambda)$ is a $2\times2$ matrix, $b(\lambda)$ is a two entries column vector and $\bar{b}(\lambda)$ is a two entries row
vector.
These coefficients are called \scatcoef{}.
From the knowledge of the scattering coefficients, it is possible to reconstruct the signal $\nfld[1,2](t)$ uniquely.

Analogously to the case of the \ac{NLSE}~\cite{Yousefi2014} we can define the \ac{NFT} continuous and discrete spectral amplitudes for the \ac{MS} as:
\begin{subequations}
\begin{align}
\cnft&=\nftb\nfta^{-1} & \eig&\in \R\\
\dnft[i]&=\nftb[][i]\nftaderiv[i]^{-1}  & \eig[{1,... n}]&\in \C \setminus \R
\end{align}
\end{subequations}
and $\nftaderiv[i]=\frac{da(\eig)}{d\eig}|_{\eig=\eig[i]}$ $\forall$ $\eig[1,....n]\in \C \setminus \R$ such
that $\nfta[i]=0$.
Although these spectral amplitudes are commonly used, it is more convenient to work
directly with the \scatcoef{} $\nfta$ and $\nftb$~\cite{HongKong}; hence, when throughout the whole manuscript we will refer to the nonlinear spectrum, we will implicitly mean the eigenvalues and the associated scattering coefficients.
The scattering coefficients are time independent and their spatial evolution is given by~\cite{Ablowitz2004a}:
\begin{subequations}
\begin{align}\label{eq:ch_transfer_function}
a(\eig,z)&=a(\eig, 0) & \bar{a}(\eig,z)&=\bar{a}(\eig,0) \\
b(\eig,z)&=b(\eig,0)e^{-4i\eig^2z} & \bar{b}(\eig,z)&=\bar{b}(\eig,0)e^{4i\eig^2z}.\label{eq:bscatcoeff}
\end{align}
\end{subequations}
In order to not overburden the notation we will drop the explicit space dependence as we did in the beginning of this section.
The fact that the \scatcoef{} are time invariant, allows computing them at an arbitrary instant of time. For example, using \eqref{eq:jost_projection_1} and the boundary Jost solutions, they can be computed at $t = +\infty$. At this instant $\jostp$ is known. Moreover, it is possible to propagate $\jostn$ from $t = -\infty$, where it is known, to $t = +\infty$ by integrating \eqref{eq:jost_projection_1}. Given the particular structure of the Jost solutions it results that the \scatcoef{} are given by:

\begin{subequations}
\begin{align}
\nfta&=\lim_{t \to +\infty}\left[\jostn[1]\jostpconj[1]^{-1}\right]\\
\nftb[1]&=\lim_{t \to +\infty} \left[\jostn[2] \jostp[2,1]^{-1}\right]\\
\nftb[2]&=\lim_{t \to +\infty} \left[\jostn[3] \jostp[3,2]^{-1}\right]
\end{align}
\end{subequations}
and using \eqref{eq:jost} and \eqref{eq:jost_projection_1} gives
\begin{subequations}
\begin{align}
\nfta&=\lim_{t \to +\infty}\left[\jostn[1]e^{i\lambda t}\right]\\
\nftb&=\begin{pmatrix}
           \nftb[1]\\
           \nftb[2]
       \end{pmatrix} =
      \lim_{t \to +\infty}\left[\begin{pmatrix}
       \jostn[2]  \\
       \jostn[3]
     \end{pmatrix}e^{-i\lambda t}\right].
\end{align}
\end{subequations}
It should be noted that, compared to the \ac{NLSE} case, there is an additional scattering coefficient $\nftb[2]$ that can be used to encode information, potentially doubling the system transmission rate.

\subsection{Inverse NFT}

The \ac{INFT} is the mathematical procedure that allows constructing a time domain  waveform starting from a given nonlinear spectrum. In our work, we have performed the \ac{INFT} at the transmitter by using an algorithm based on the \ac{DT}~\cite{matveev1991darboux}.
The \ac{DT} is a natural candidate to build time domain signals, especially when the information is encoded only in the discrete nonlinear spectrum. The method consists in adding iteratively discrete eigenvalues to the nonlinear spectrum while simultaneously updating the signal in time domain. The \ac{INFT} based on the \ac{DT} for eigenvalue communications was proposed in~\cite{Yousefi2014a}.
In our work we have used the \ac{DT} for the \ac{MS} derived
by Wright~\cite{Wright2003}. We summarize here how the \ac{DT} for the \ac{MS} works.

\begin{figure}[!t]
  \centering
  \includegraphics[width=\columnwidth]{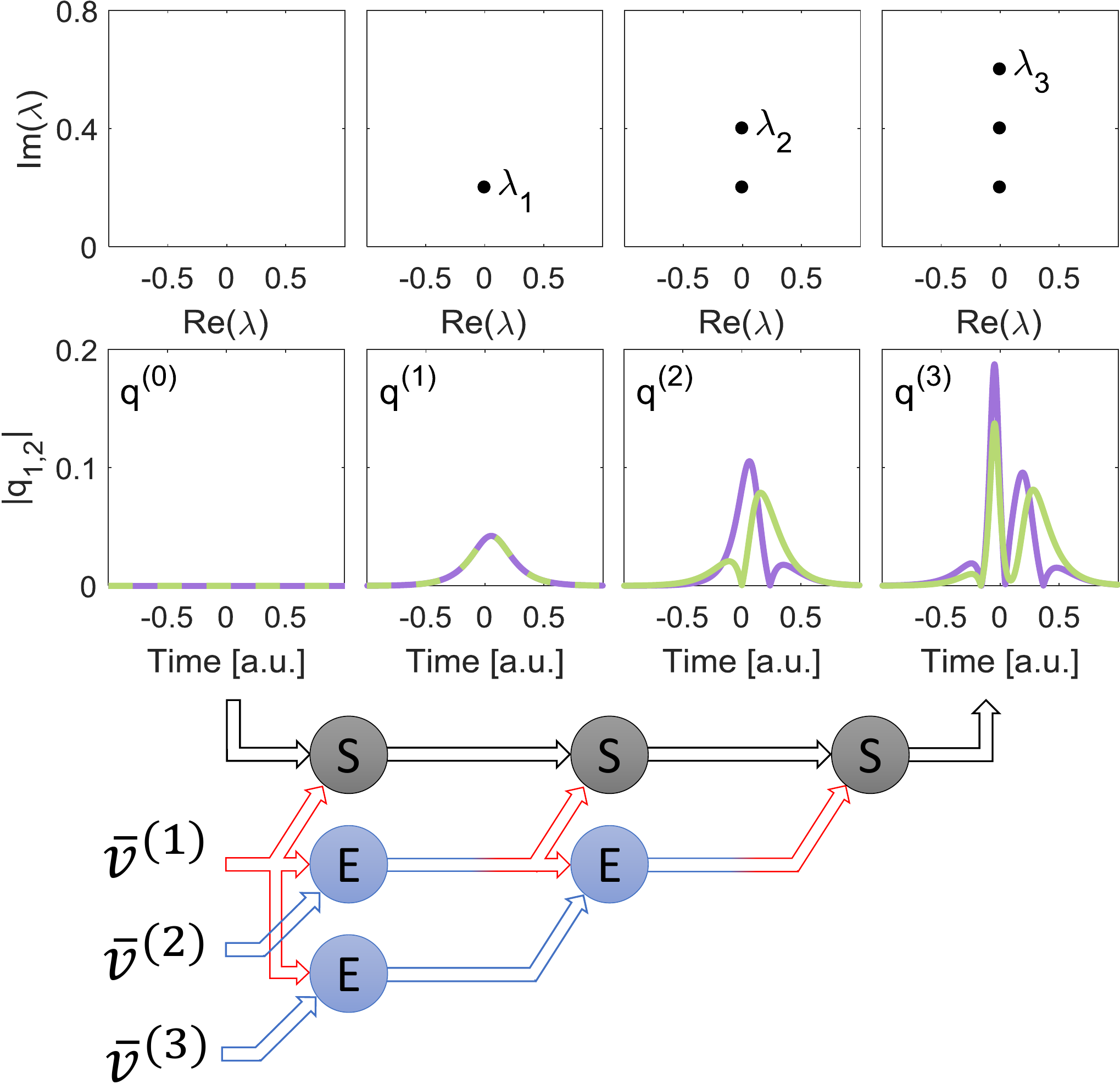}
  \caption{Schematic of the \ac{DT}. The S-node is the signal update operation corresponding to \eqref{eq:darboux_signal_update} and the E-node is the eigenvector
  update operation corresponding to \eqref{eq:darboux_eigenvectors_update}. At the step $i=1,2,3$ the auxiliary solution $\auxsol$\textsuperscript{(i)} for $\eig = \eig[i]$ (red arrow) modifies the signal $q_j$\textsuperscript{(i-1)} and all the other auxiliary solutions (blue arrows). The seed null signal $q_j$\textsuperscript{(0)},~$j=1,2$ entering the first S-node is transformed after each step in such a way that its discrete spectrum has a new eigenvalue added as shown in the four insets in upper part of the figure.}
  \label{fig:darbouxlower}
\end{figure}

Let $\vecv$ be a column vector solution of the \ac{MZSP}  spectral problem \eqref{eq:zsp} associated with the \ac{MS}  for the
signal $\nfld(\nttm)$ and the eigenvalue $\eig$, then according to~\cite{Wright2003} a new solution of \eqref{eq:zsp}, $\hat{\vecv}$, is given by the following equation:
 \begin{align}\label{eq:darboux_eigenvectors_update}
 \hat{\vecv}=\left(\lambda \mathcal{\matr{I}}_3-\matr{G}_0\right)\vecv
 \end{align}
where $\mathcal{\matr{I}}_3$ is the $3\times3$ identity matrix, $\matr{G}_0=\auxsolmat \matr{M}_0\auxsolmat^{-1}$ with
\begin{align}\label{eq:auxsol_matrix}
 \auxsolmat= \begin{pmatrix}
       \auxsol_1 & \auxsol_2^* & \auxsol_3^* \\
       \auxsol_2 & -\auxsol_1^*  & 0 \\
       \auxsol_3 & 0 & -\auxsol_1^*
     \end{pmatrix}
\end{align}
where the matrix $\matr{M}_0=diag(\lambda_0,~\lambda_0^*,~\lambda_0^*)$ and $ \auxsol=(\auxsol_1,~\auxsol_2,~\auxsol_3)^T$ is a solution of \eqref{eq:zsp} for the seed signal $q_j,~j=1,2$ and a fixed value of $\eig = \eig[0]$. The \ac{DT} gives the new signal waveforms in time domain for both polarizations \mbox{$\hat{q}_j,~j=1,2$} as a function of the old signals $q_j$, of the auxiliary solution $\auxsol$ and of the new eigenvalue $\lambda_0$ we want to add to the nonlinear spectrum:
\begin{align}\label{eq:darboux_signal_update}
 \hat{q}_j=q_j+2i(\lambda_0^*-\lambda_0)\frac{u_j^*}{1+\sum_{s=1}^2|u_s|^2} \ \ (j=1,2)
 \end{align}
where $u_j=\auxsol_{j+1}/\auxsol_1$.

Starting from the "vacuum" solution $\nfld[j](t)=0,~j=1,2$, the procedure sketched in \figurename~\ref{fig:darbouxlower} can be repeated iteratively to generate the dual polarization time domain  signal associated with a nonlinear spectrum containing an arbitrary large number of discrete eigenvalues.

The generic auxiliary solution $\auxsol$\textsuperscript{(k)} that satisfies the \ac{MZSP} for the eigenvalue $\lambda_k$ reads: $\auxsol$\textsuperscript{(k)} $=(A$\textsuperscript{(k)}$e^{-i\lambda_k t},~B$\textsuperscript{(k)}$e^{i\lambda_k t},~C$\textsuperscript{(k)}$e^{i\lambda_k t})^T$ (being $\{A$\textsuperscript{(k)}$,~B$\textsuperscript{(k)}$,~C$\textsuperscript{(k)}$\}$ some initialization constants). Hence after adding $i$ eigenvalues the auxiliary solutions are modified according to the following chain of matrix multiplications
\stackMath
 \begin{eqnarray}
\begin{pmatrix}
       \stackon[-3pt]{\auxsol}{\hat{}}_1^{(k)} \\
       \stackon[-3pt]{\auxsol}{\hat{}}_2^{(k)} \\
       \stackon[-3pt]{\auxsol}{\hat{}}_3^{(k)}
     \end{pmatrix}=
     \left(\lambda_k\mathcal{\matr{I}}_3-\matr{G}_{0i-1}\right)...\left(\lambda_k\mathcal{\matr{I}}_3-\matr{G}_{01}\right)
     \begin{pmatrix}
      A^{(k)}e^{-i\lambda_k t} \\
       B^{(k)}e^{i\lambda_k t} \\
      C^{(k)}e^{i\lambda_k t}
     \end{pmatrix}
     \end{eqnarray}
where the $\matr{G}_{0i}$ matrices are evaluated as functions of the $i$-th auxiliary
solution $\auxsol^{(i)}$ evaluated  after $i-1$ Darboux transformations, see also the scheme depicted in
\figurename~\ref{fig:darbouxlower}.
The initialization constants $\{A^{(k)}, B^{(k)}, C^{(k)}\}$
have been set respectively equal to $\left\{1,-b_{1}(\lambda_k),-b_{2}(\lambda_k)\right\}$,  in order to obtain the correct spectrum after performing \ac{NFT} and \ac{INFT} in sequence as in the scalar case~\cite{Aref2016c}.
$b_{1}(\lambda_k)$ and $b_{2}(\lambda_k)$ are the \scatcoef{} that we want to associate to the eigenvalue $\lambda_k$ for the two polarization components respectively.


\section{\ac{DP-NFDM} system}\label{sec:nfdm_system}
In this section, the basic structure of a \ac{DP-NFDM} system using the discrete spectrum will be described. The \ac{DSP} chain will be introduced first and then the experimental setup.

\subsection{Transmitter and receiver digital signal processing}

\begin{figure}[!t]
  \centering
  \includegraphics[width=\columnwidth]{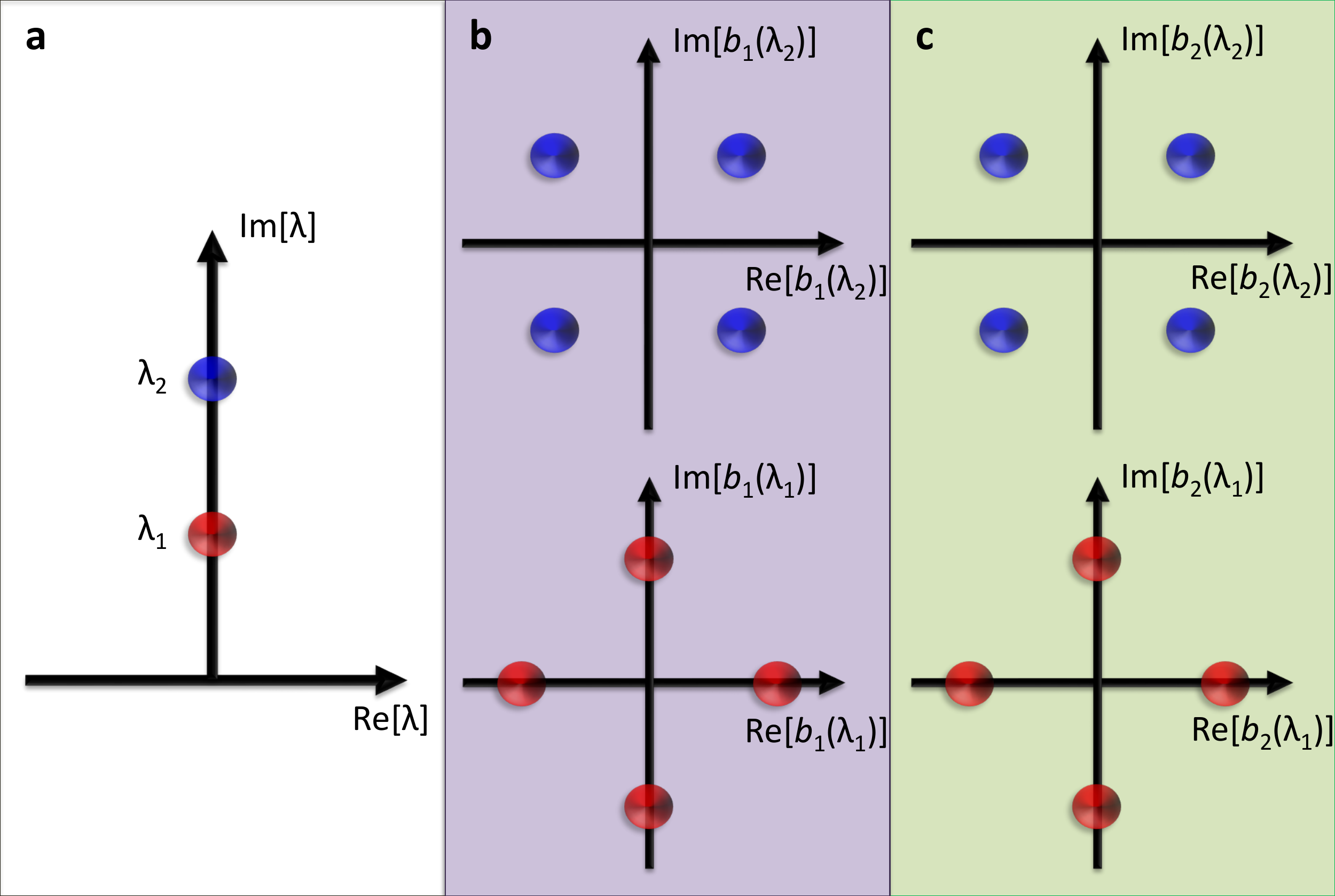}
  \caption{Ideal normalized constellations are illustrated schematically: in (\textbf{a}) the discrete eigenvalues $\lambda_1 = i 0.3$ and $\lambda_2~=~i 0.6$ are depicted.
  The scattering coefficients $\nftb[1,2][i],~i=1,2$, associated with the two orthogonal polarization components of the signal, are shown in (\textbf{b}) and (\textbf{c}) respectively. Polarization~1 and Polarization~2 on a violet and green background respectively. The scattering coefficients associated with $\eig[1]$ are chosen from a \ac{QPSK} constellation  of radius 5 and rotated by $\pi/4$ while those associated with $\eig[2]$ from a \ac{QPSK} constellation  and radius 0.14.
  }
  \label{fig:idealconst}
\end{figure}

\begin{figure*}[!t]
  \centering
  \includegraphics[width=\textwidth]{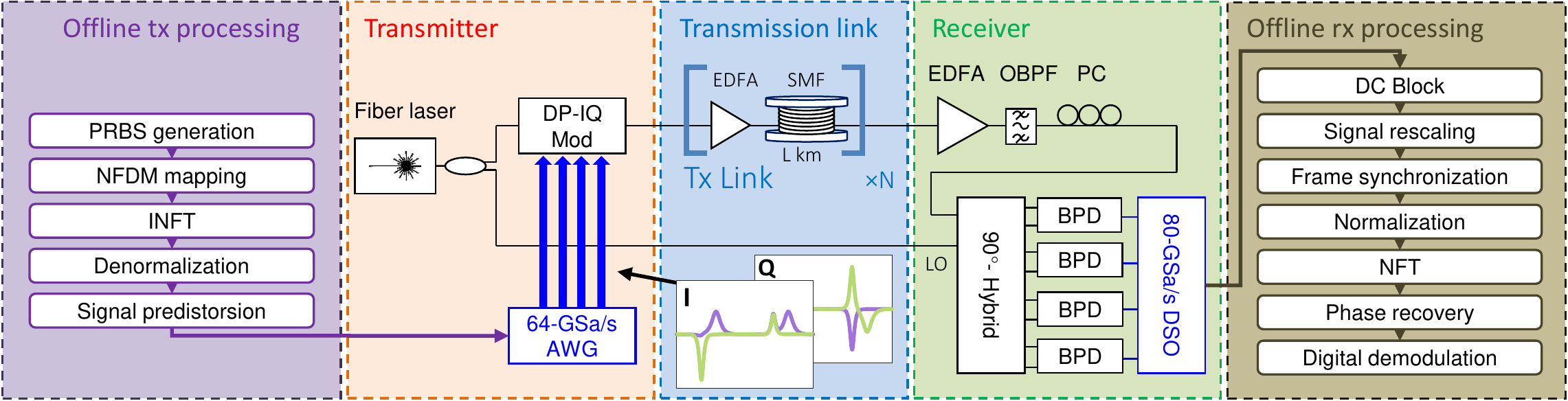}
  \caption{Experimental setup with transmitter and receiver \acs{DSP} chain. Abbreviations not defined in the main text: \ac{BPD}, \ac{DC}.}
  \label{fig:setup}
\end{figure*}

At the transmitter the data bits are mapped to the \scatcoef{} pairs $\{\nftb[1][i],~\nftb[2][i]\}$ for $i=1,2$ where the pair of eigenvalues $\{\eig[1] = i0.3,~\eig[2] = i0.6\}$ is used for each symbol. We will refer to these set of coefficients and equivalently to the associated time domain waveform as a \ac{DP-NFDM}~symbol. The \scatcoef{} associated with the first eigenvalue can assume values drawn from a \ac{QPSK} constellation of radius 5 and rotated by $\pi/4$ while those associated with the second eigenvalue are drawn from a \ac{QPSK} constellation of radius 0.14 as shown in \figurename~\ref{fig:idealconst}. This particular structure of the constellations was chosen to reduce the \ac{PAPR} of the signal at the transmitter, in order to limit the performance losses due to the limited resolution of the \ac{DAC} and due to the nonlinear characteristic of \acp{MZM} and electrical amplifiers (See Supplementary Material for a detailed explanation).
The waveform associated to each \ac{DP-NFDM} symbol is generated using the \ac{DT} described in the previous section followed by the denormalization as in \eqref{eq:normalization}  with normalization parameter
$T_0 = $47 ps. This choice of $T_0$ allows fitting the waveform in a time window of 1~ns (1~GBd) with enough time guard band among successive \ac{DP-NFDM} symbols to satisfy the vanishing boundary conditions required to correctly compute the \ac{NFT}. The power $P_{tx}$ of the digital signal thus obtained will later be used to set the power of the corresponding transmitted optical signal.

The channel is assumed to be a link of standard \ac{SMF} with \ac{EDFA}
lumped amplification as in the experiment. In order to take into account the presence of the
losses, the \ac{LPA} approximation is used in the normalization and
denormalization steps of the waveform before computing the \ac{NFT} and after
computing the \ac{INFT} respectively.

\begin{figure}[!b]
  \includegraphics[width=0.95\columnwidth]{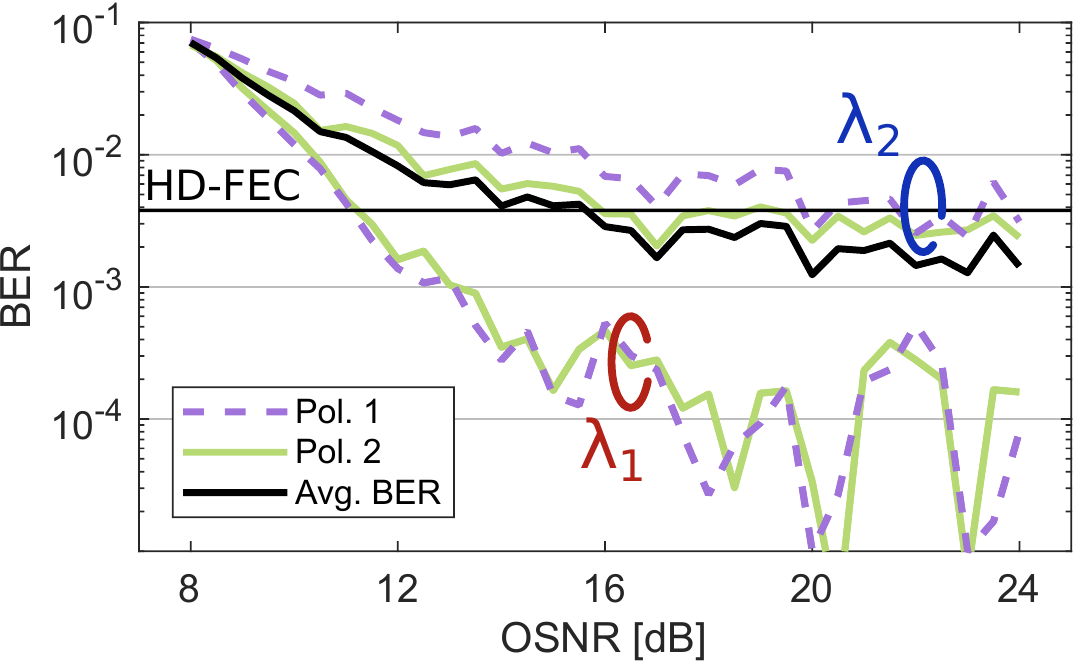}
  \caption{System performance in terms of \ac{BER} as a function of the \acs{OSNR} in a back-to-back configuration. The \ac{BER} of the individual constellations are shown by the violet (Polarization 1) and green (Polarization 2) curves and are grouped per eigenvalue ($\eig[1] = i0.3,~\eig[2] = i0.6$). The black curve represents the average \ac{BER} over the 4 constellations. }
  \label{fig:b2bperfomance}
\end{figure}
At the receiver, the digital signal output by the \ac{DSO} is first rescaled so that its power is $P_{tx}$ (the power of the transmitted optical signal). Then an ideal rectangular filter with bandwidth equal to the 99\% power bandwidth of the signal is used to filter the out of band noise.
At this point, cross-correlation-based
frame synchronization using training sequences is performed in order
to optimally align the \ac{DP-NFDM}~symbol to the processing window.
For each \ac{DP-NFDM} symbol, first the eigenvalues are
located using the Newton-Raphson search method employing the one-directional trapezoidal method and then the coefficients $\nftb[1,2][i]$
are computed on the found eigenvalues using the forward-backward trapezoidal  method (see Supplementary Material for more details). The homodyne configuration of the receiver allows not having a frequency offset between the transmitter laser and the coherent receiver \ac{LO}, but given the non-zero combined linewidth of the two lasers (\textasciitilde 1kHz) their coherence length is limited to about 90~km. This implies that the received constellations are affected by phase noise when the transmission distance exceeds the coherence length of the laser, causing errors in the detection of the symbols. The phase noise is removed by applying the blind phase search algorithm \cite{Pfau:09} in the \ac{NFT} domain to each constellation individually.
Finally, the \scatcoef{} are rotated back to remove the phase factor acquired during the transmission (\eqref{eq:bscatcoeff}) and the decision on the symbols is taken using a minimum Euclidean distance decisor over the \scatcoef{}.

\begin{figure*}[!ht]
  \centering
  \includegraphics[width=.366\textwidth]{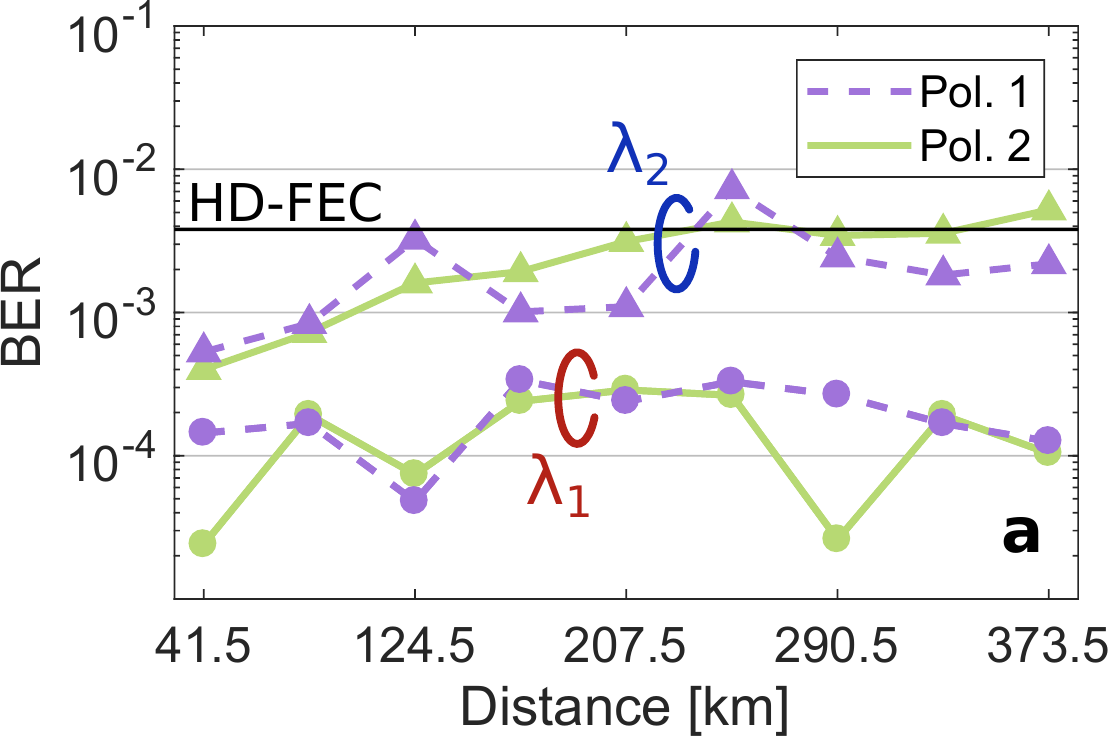}
  \includegraphics[width=.3095\textwidth]{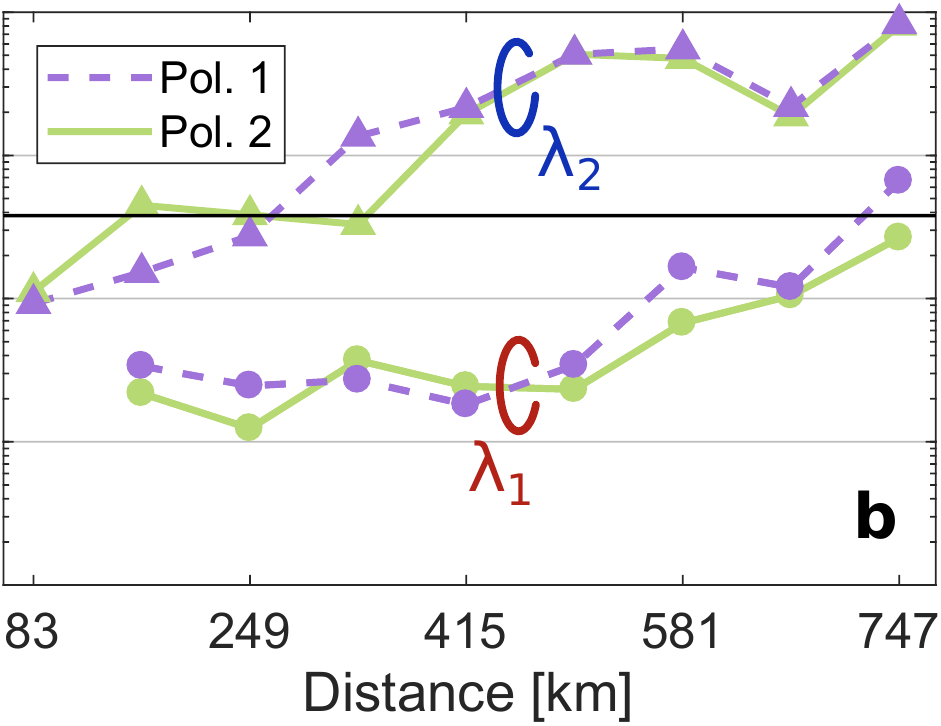}
  \includegraphics[width=.315\textwidth]{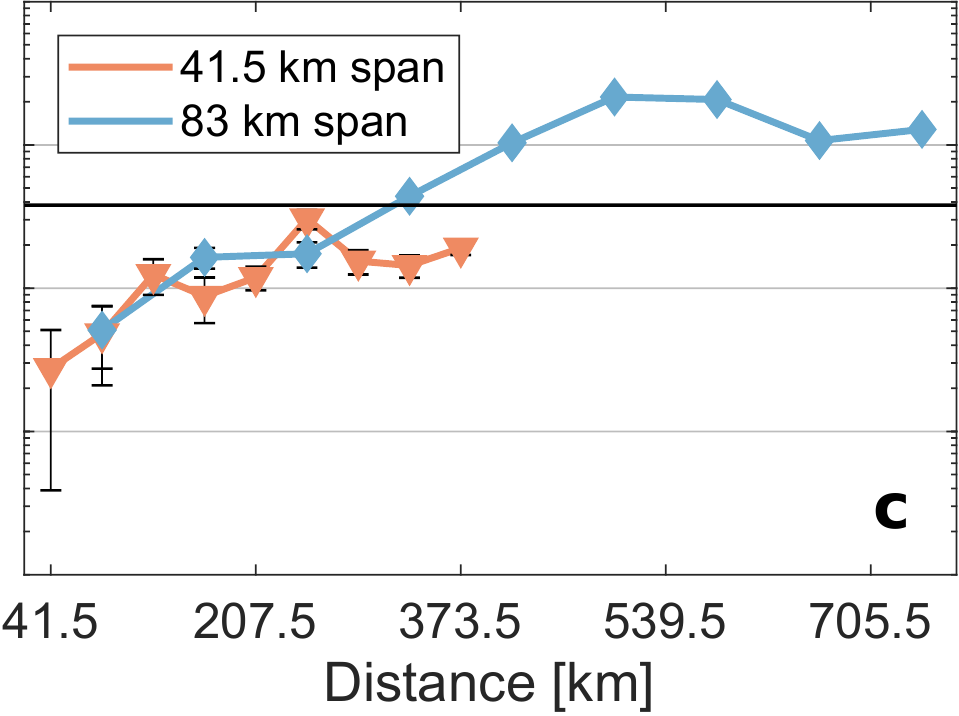}
  \caption{System performance in terms of \ac{BER} as a function of the transmission distance for the 4 individual constellations for \mbox{$L=41.5$~km} (\textbf{a}) and $L=83$~km (\textbf{b}) spans. The violet (Polarization 1) and green (Polarization 2) curves are grouped per eigenvalue ($\eig[1] = i0.3,~\eig[2] = i0.6$). (\textbf{c}) Comparison of the average \ac{BER} versus transmission distance between links of the two different span lengths. The error bars represent the standard deviation over 5 processed blocks of $10^5$ \ac{DP-NFDM}~symbols.}
  \label{fig:performance}
\end{figure*}

\subsection{Experimental setup}\label{experimental-setup}

The experimental setup and the block diagrams of the \ac{DSP} are depicted  in \figurename~\ref{fig:setup}.
At the
transmitter a \ac{FL} with sub-kHz
linewidth is modulated using an integrated dual polarization I/Q
modulator driven by an \ac{AWG} with 20~GHz analog
bandwidth and 64~GSa/s. Before uploading it to the \ac{AWG}, the signal
generated by the \ac{INFT} is pre-distorted using the ideal inverse
transfer function of the \ac{MZM} ($\textnormal{asin}(\cdot)$). This pre-distortion is
required in order to have a good trade-off between \ac{SNR} at the output of
the \ac{MZM} and signal distortions caused by its nonlinear transfer function. Nonetheless, given the still high \ac{PAPR} of the optimized waveform considered (see Supplementary Material), this
pre-distortion is not optimal and advanced methods can be employed to
improve further the quality of the transmitted signal~\cite{Le2017}.
The channel is a fiber link composed of
up to 9 spans of \ac{SMF} fiber with dispersion \mbox{$D =
17.5$~ps/nm${\cdot}$km}, nonlinear coefficient \mbox{$\gamma$ = 1.25~W$^{-1}$km$^{-1}$}, attenuation \mbox{$\alpha$ = 0.195~dB/km} and \ac{PMD} coefficient < 0.1 ps km$^{-1/2}$.
Two different span lengths of $L=41.5$ and $L=83$~km were employed.
Considering these channel parameters, the complex baseband signal generated by the \ac{INFT} with \ac{LPA} and denormalized has the following properties: 99\% of its power contained within a bandwidth $W$ of 12.7~GHz, a \ac{PAPR} of 9.49~dB and an average power $P_{tx}$ of 5.30~dBm and 7.70~dBm for the span lengths $L$ of 41.5 and 83~km, respectively. Given these channel and signal parameters we have that the soliton period, defined as $(\pi/2) L_d$, with $L_d = (W^{}|\dispersion|)^{-1}$ the dispersion length \cite{hasegawa1995solitons,Turitsyn2017}, is 436 km. Being this  much larger than the typical birefringence correlation length, which is on a scale of few tens of meters \cite{menyuk2006interaction}, guarantees the applicability of the Manakov averaged model.

In order to properly match the transmitted signal to the channel, the gain of the \ac{EDFA} at the transmitter is tuned in such a way to set the power of the optical signal to $P_{tx}$.
The optical signal is then transmitted through the channel.

At the receiver, the signal is first sent through a 0.9~nm \ac{OBPF} and then a \ac{PC} was used to manually align
the polarization of the signal to the optical front-end. The use of the \ac{PC} was required to avoid the use of polarization tracking algorithms for the \ac{NFT} signals, which were not available at the time of the experiment. In the future it could be possible to use modulation independent polarization tracking algorithms, as an example using independent components analysis \cite{nabavi2015demultiplexing}. The signal is then detected by using a standard coherent receiver (33 GHz analog bandwidth, 80~GSa/s), in a homodyne configuration where the transmitter laser is used as \ac{LO}. The acquired digital signal consisting of 5 blocks of $10^5$ DP-NFDM symbols is then fed to the receiver \ac{DSP} chain described previously.


\section{Experimental results}\label{sec:experimental_results}
\label{experimental-results}

The system was initially tested in a \ac{B2B} configuration, where the transmitter output has been directly connected to the receiver, in order to obtain the best performance achievable by the system in the sole presence of the intrinsic transceiver distortions (e.g. transmitter front-end distortions, detectors noise, etc.) and added \ac{AWGN} as commonly done for linear coherent systems. The \ac{OSNR} was swept by varying the noise power added to the signal at the receiver input. The adopted metric for measuring the performances allows a direct comparison with standard coherent transmission systems. The \ac{OSNR} range considered is the region of interest where the system performance is around the HD-FEC threshold.  The measured average \ac{BER}
is shown in \figurename~\ref{fig:b2bperfomance}.
A visible effect is the fact that the \ac{BER} is not the same for the four different
constellations, but it is worse for the two constellations associated with
the eigenvalue with a higher imaginary part.
This effect can be related to the dependency of the noise variance of both the eigenvalues and the corresponding scattering coefficients on the imaginary part of the eigenvalues themselves \cite{zhang2015gaussian,Zhang1,Zhang2,Hari1,Hari2}; an analysis of the noise distribution for the various eigenvalues and scattering coefficients is provided in the Supplementary Material.

\begin{figure}[!b]
  \centering
  \includegraphics[width=\columnwidth]{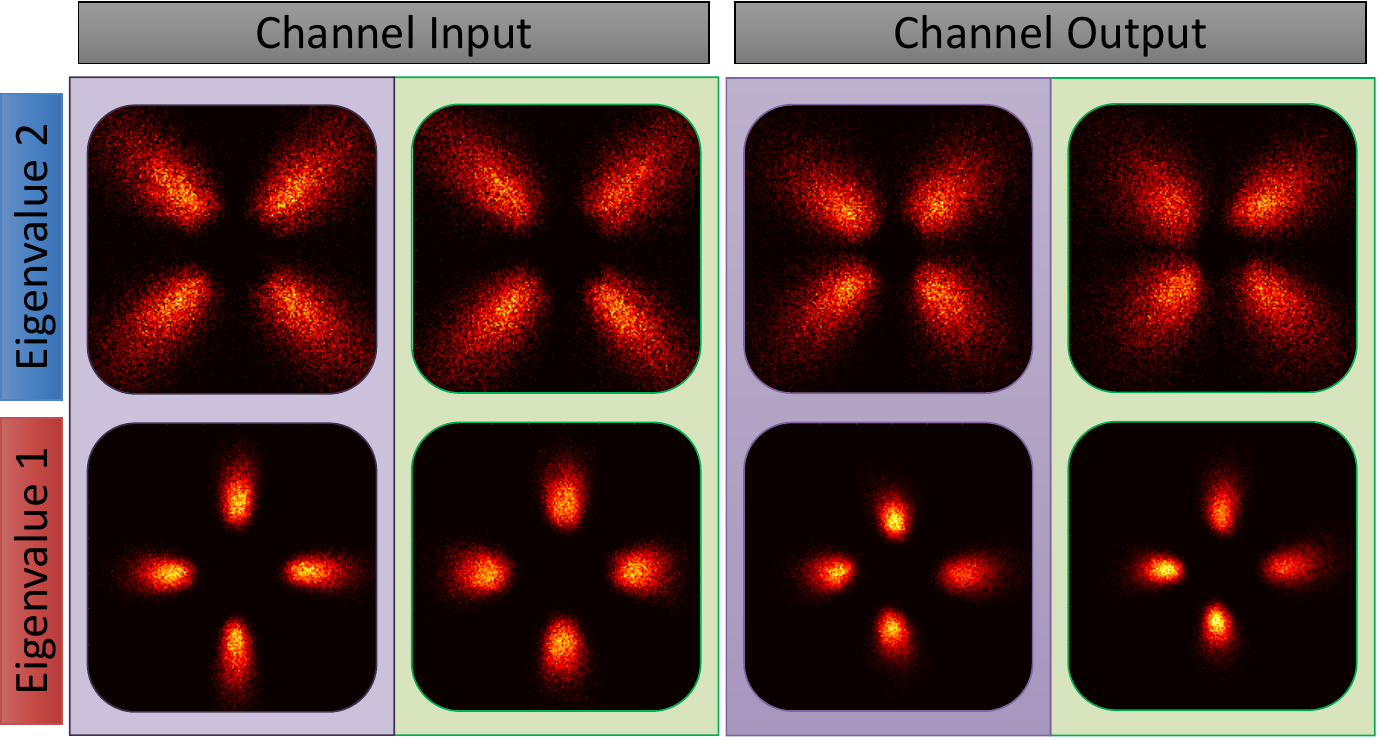}
  \caption{The four experimental constellations of the scattering coefficients $\nftb[1,2][i],~i=1,2$ associated with the two eigenvalues ($\eig[1] = i0.3,~\eig[2] = i0.6$) are shown at the transmitter side (left) and after 373.5~km transmission with 41.5~km spans (right). Polarization~1 and Polarization~2 on a violet and green background respectively.}
  \label{fig:noisy_constellations}
\end{figure}

In order to demonstrate fiber transmission with the proposed system, a transmission was
performed over a link of $N$~spans of \ac{SMF}  with span length $L=41.5$ and $L=83$~km. The
performance in terms of \ac{BER} as a function of the transmission distance
is shown in \figurename~\ref{fig:performance}~(a-b) for the four different constellations. The difference
in performance in the two eigenvalues appears in this case too. This can also be seen from the constellation
plots after 373.5~km in \figurename~\ref{fig:noisy_constellations}, where the two constellations associated with $\eig[2]$ are
sensibly more degraded than those related to $\eig[1]$, which are still well
defined. Similar performance can instead be seen in the two different
polarizations of the same eigenvalue.

Finally in \figurename~\ref{fig:performance}~(c) we compare the average \ac{BER} for the two span lengths
used in the test, in order to check the impact of the \ac{LPA} approximation.
Worse performances are expected when longer links are used being $\gamma_{eff}$ given in \eqref{eq:LPA} farther from  the real $\gamma$ of the fiber in this case.
The \ac{BER} curve for the 41.5~km span contains two outlier points at 124.5~km and 249~km
that are slightly worse than the general trend of the curve.
This is believed to be caused by instabilities in the setup when the related experimental traces were acquired, in particular an incorrect alignment of the polarization to the receiver due to a drift in the polarization state of the received signal. Besides these two points, the rest of the 41.5~km curve lies
under the one for the 83~km spans, confirming that the use of longer
spans adds a slight degradation in the performance of the system. The
maximum reach of the system achieved with \ac{BER} under the \ac{HD-FEC} threshold
is 373.5~km using 41.5~km spans and 249~km with spans of 83~km.
It should be noted that in our experimental setup the \ac{PMD} effect was not compensated for.
However, for the transmission lengths and \ac{PMD} values of the standard \ac{SMF} employed, the accumulated differential group delay is negligible if compared with the pulse duration \cite{AgrawalBook}. The impact of \ac{PMD} is therefore not expected to have had a major impact on the results shown. New approaches have been developed to compensate for PMD  effects in linear transmission systems \cite{kikuchi2011analyses} and a recent work has shown in simulations that for a  \ac{DP-NFDM} system employing the continuous spectrum, \ac{PMD} effects could be compensated in the nonlinear domain by using a linear equalizer \cite{Goossens:17}. Similar techniques may be applied to discrete \ac{DP-NFDM} systems.


\section{Conclusions}\label{sec:nfdm_summary}

We have demonstrated experimentally, for the first time to the best of our knowledge, an eigenvalue-based optical communication system employing two orthogonal modes of polarization. We encoded 8 bits/\ac{DP-NFDM} symbol and demonstrated transmission up to 373.5~km. Furthermore, we have shown that a powerful, but rather abstract mathematical technique, the Darboux transformation, can have indeed far-reaching impact in applied nonlinear optics namely in fiber-based telecommunication systems. Our results pave the way towards doubling the information rate of \ac{NFT}-based fiber optics communication systems. Although more research work needs to be done in this direction, by demonstrating the possibility of using dual polarization \ac{NFT} channels, we have indeed successfully met one of the key challenges that were explicitly highlighted in a recent review of this research field~\cite{Turitsyn2017} as necessary steps in order to bring eigenvalue communication from a pioneering stage to be a working infrastructure for optical communications in the real-world. Furthermore, the demonstration of the polarization division multiplexing is a significant step forward towards a fair comparison of the \ac{NFT}-based channels with the currently used linear ones where polarization division multiplexing is an established practice.
\section*{Funding Information}
This work was supported by the Marie Curie Actions through ICONE Project (no. 608099), the DNRF Research Centre of Excellence, SPOC (no. DNRF123) and the Villum foundation.

\section*{Acknowledgments}

The authors thank S.K. Turitsyn, M. Kamalian, Y. Prylepskiy, R. T. Jones and J. Diniz for the stimulating discussions, constant encouragement and careful reading of the manuscript and the anonymous reviewers for their constructive criticism and relevant suggestions that improved the quality of the manuscript.

\vspace{0.4cm}
\noindent
See Supplement 1 for supporting content.


\bibliography{bibliography}

\end{document}